\documentclass[12pt,preprint]{aastex}

\usepackage{color}

\usepackage{graphicx}
\usepackage{rotating}
\shorttitle{}
\shortauthors{Nesvorn\'y et al.}

\begin{document}
\baselineskip 19.pt
\title{Radial Distribution of Distant Trans-Neptunian Objects Points to Sun's Formation in a 
Stellar Cluster}
\author{David Nesvorn\'y$^{1,*}$, Pedro Bernardinelli$^{2}$, David Vokrouhlick\'y$^3$, Konstantin Batygin$^4$} 
\affil{(1) Department of Space Studies, 
Southwest Research Institute, 1050 Walnut St., Suite 300,  Boulder, CO 80302, United States}
\affil{(2) DIRAC Institute, University of Washington, 3910 15th Ave NE, Seattle, WA 98195-0002, United States}
\affil{(3) Institute of Astronomy, Charles University, V Hole\v{s}ovi\v{c}k\'ach 2, CZ–18000 Prague
  8, Czech Republic}
\affil{(4) California Institute of Technology, 1200 East California Boulevard, Pasadena, California 91125,
United States}
\affil{*e-mail:davidn@boulder.swri.edu}

\begin{abstract}
The Scattered Disk Objects (SDOs) are a population of trans-Neptunian bodies with semimajor axes 
$50< a \lesssim 1000$ au and perihelion distances $q \gtrsim 30$ au. The detached SDOs with 
orbits beyond the reach of Neptune (roughly $q>35$~au) are of special interest here as an important 
constraint on the early evolution of the outer Solar System. The semimajor axis profile of detached 
SDOs at 50--500~au, as characterized from the Dark Energy Survey (DES), is radially extended,
but previous dynamical models of Neptune's early migration produce a relatively compact profile.  
This problem is most likely related to Sun's birth environment in a stellar cluster. 
We perform new dynamical simulations that account for cluster effects and show that the orbital 
distribution of SDOs can be explained if a particularly close stellar encounter occurred early on 
(e.g., M dwarf with the mass $\simeq 0.2$ $M_\odot$ approaching the Sun at $\simeq 200$ au). For such 
an encounter to happen with a reasonably high probability the Sun must have formed in a stellar 
cluster with $\eta T \gtrsim 10^4$ Myr pc$^{-3}$, where $\eta$ is the stellar number density and
$T$ is the Sun's residence time in the cluster.     
\end{abstract}

\section{Introduction}
The dynamical structure of the Kuiper Belt can be used as a clue to the formation and evolution 
of the Solar System, planetary systems in general, and Neptune's early orbital history in particular. 
The exact nature of Neptune's orbital migration has been the subject of considerable research 
(see Morbidelli \& Nesvorn\'y 2020 and Gladman \& Volk 2021 for recent reviews). The problem is best addressed by forward 
modeling where different initial conditions and Neptune's orbital evolutions are tested, and the 
model predictions are compared to observations. Initial studies envisioned dynamical models where 
Neptune maintained a very low orbital eccentricity, comparable to the present $e_{\rm N} \simeq 0.01$, 
during its early migration (e.g., Malhotra 1993, 1995; Gomes 2003; Hahn \& Malhotra 2005). A 
giant-planet instability model was later proposed to explain the somewhat excited orbits of the 
outer planets (Tsiganis et al. 2005). In the original instability model, Neptune was scattered 
to a highly eccentric orbit ($e_{\rm N} \gtrsim 0.2$) that briefly overlapped with the Kuiper Belt 
(Levison et al. 2008). 

Different arguments have been advanced to rule out specific migration/instability regimes (e.g., 
Batygin et al. 2011; Dawson \& Murray-Clay 2012). For example, the high-eccentricity instability 
model with rapid (e-fold $\tau \sim 1$ Myr) circularization of Neptune's orbit does not reproduce the wide 
inclination distribution of Kuiper belt objects (KBOs), because there is not enough time to dynamically 
excite orbits in this model (Volk \& Malhotra 2011, Nesvorn\'y 2015a). The migration models with a very low 
eccentricity of Neptune ($e_{\rm N} \lesssim 0.03$; Volk \& Malhotra 2019) do not explain KBOs 
with $50<a<60$ au, perihelion distances $q>35$ au, and $i<10^\circ$ (Nesvorn\'y 2021). Most 
modern studies therefore considered a mild version of the instability with $e_{\rm N} \simeq 
0.03$--0.1 and $\tau \sim 10$ Myr (Nesvorn\'y \& Morbidelli 2012; Kaib \& Sheppard 2016; 
Deienno et al. 2017, 2018; Lawler et al. 2019; Clement et al. 2021).

\begin{figure}
\epsscale{0.7}
%\plotone{intro_fig.JPG}
\plotone{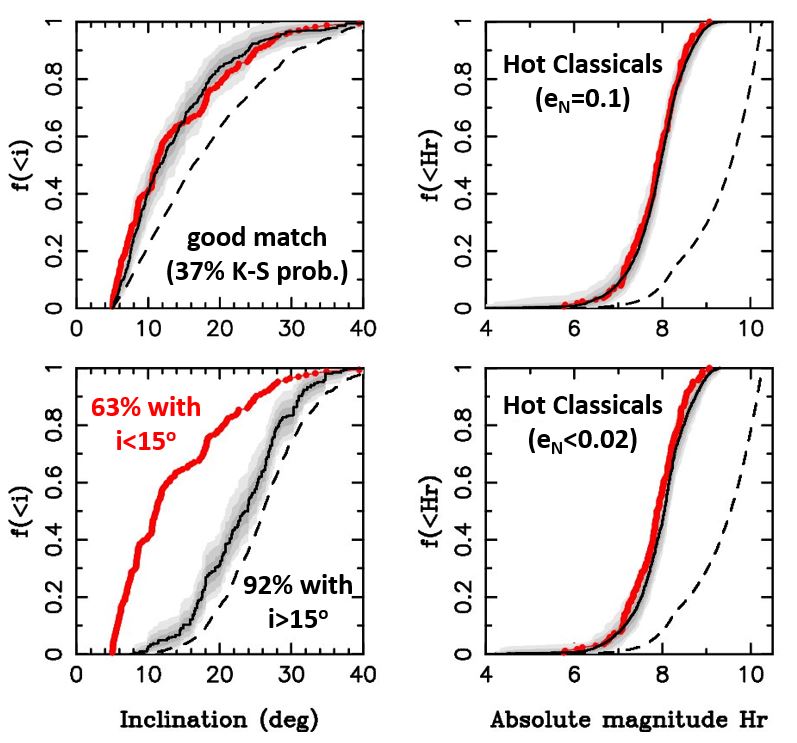}
% bottom is zero2, up is galaxy
\caption{A comparison between the biased model (black lines) and OSSOS observations (red dots) of Hot 
Classicals ($40<a<47$ au, $q>36$ au, $i>5^\circ$; Nesvorn\'y et al. 2020). The intrinsic model distributions 
are shown as dashed lines. The shaded areas are 1$\sigma$ (bold gray), 2$\sigma$ (medium) and 3$\sigma$ 
(light gray) envelopes. We used the model results and generated 10,000 random samples with 164 bodies each (the sample size identical to the 
number of OSSOS detections). The samples were biased with the OSSOS simulator (Lawler et al. 2018).
We identified envelopes containing 68.3\% (1$\sigma$), 95.5\% (2$\sigma$) and 99.7\% (3$\sigma$) of samples 
and plotted them here. The bottom panels show a model with the low-eccentricity migration of 
Neptune ($e_{\rm N} \simeq 0.01$). In this case, orbits are decoupled from Neptune by the Kozai resonance 
(Kozai 1962) and this creates a specific inclination distribution with very few 
orbits below $15^\circ$ (Nesvorn\'y 2021). The upper panels show a successful model where Neptune's 
eccentricity was excited to $e_{\rm N}=0.1$ when Neptune reached $\simeq 28$ au, and slowly damped
afterwards. The K-S test applied to the inclination distributions from the successful and unsuccessful 
cases gives 37\% and $<10^{-10}$ probabilities, respectively, that the (biased) model and observed 
distributions are drawn from the same underlying distribution. Figure adapted from Nesvorn\'y et 
al. (2020).}
\label{intro}
\end{figure}

In the previous work, we developed dynamical models for resonant and dynamically hot KBOs 
(Nesvorn\'y \& Vokrouhlick\'y 2016), dynamically cold KBOs (Nesvorn\'y 2015b) and SDOs 
(Kaib \& Sheppard 2016, Nesvorn\'y et al. 2016).
%Centaurs (Nesvorn\'y et al. 2019) and comets (Nesvorn\'y et al. 2017, Vokrouhlick\'y et al. 2019). 
The newest of these models were constrained by the Outer Solar System Origins Survey (OSSOS; 
Bannister et al. 2018) observations (Nesvorn\'y et al. 2020). 
For example, Fig. \ref{intro} compares a successful dynamical 
simulation with a model where Neptune was assumed to have migrated on a low-eccentricity orbit 
($e_{\rm N} \simeq 0.01$). The results indicate that Neptune's migration was long-range (from 
$\lesssim 25$ au to 30 au), slow ($\tau \gtrsim 10$ Myr) and grainy (due to scattering encounters 
with Pluto-sized objects), and that Neptune's eccentricity was excited to $e_{\rm N} \simeq 0.03$--0.1 when 
Neptune reached $\simeq 28$ au (probably due to an encounter with a planetary-class object). 

Here we consider SDOs. SDOs can be divided into {\it scattering} and {\it detached} populations. Gladman 
et al. (2008) defined scattering SDOs as objects that are dynamically coupled to Neptune (semimajor axis 
change $\Delta a > 1.5$ au in a 10-Myr long integration window, plus $a>47.8$ au), and detached SDOs as 
those that are not coupled (non-resonant, $\Delta a < 1.5$ au, and $e>0.24$ to avoid classical KBOs). 
Here we adopt a simpler definition. We avoid orbits with $a<60$ au because we do not want to mix 
arguments about the formation of distant/detached SDOs with those related to capture of bodies into 
the strong 5:2 resonance at $a\simeq55.5$ au (Gladman et al. 2012). The objects with $60<a<500$ au 
are separated to those with the perihelion distance $q=a(1-e)<35$ au (our scattering SDOs) and $q>35$ 
au (our detached SDOs). There is a good overlap with the definition of Gladman et al. (2008) because 
SDOs with $q<35$ au are typically scattered by Neptune while the ones with $q>35$ au are not. 
Importantly, we apply the same (our) definition to both the model and observed populations -- this 
allows us to accurately compare the two.   

\begin{figure}
\epsscale{0.7}
%\plotone{galaxy.eps}
\plotone{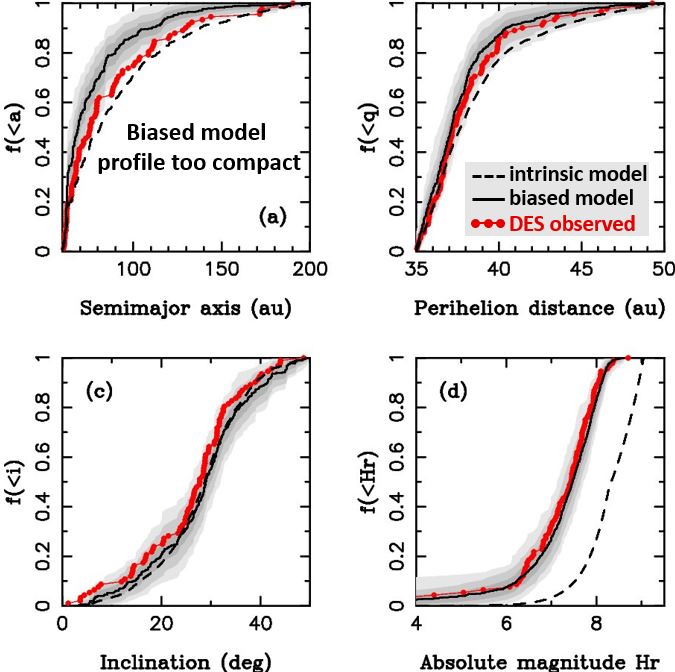}
%scatter7
\caption{The problem with detached SDOs. The plot shows a comparison between the biased model (black line) 
from our Galaxy simulation and Dark Energy Survey (DES) observations (red lines) of detached SDOs. 
See Sect. 2 for the description of model parameters and DES (Bernardinelli et al. 2022). The intrinsic 
model distributions are shown as dashed lines. See the caption of Fig. \ref{intro} for the meaning of shaded 
areas. The Kolmogorov-Smirnov (K-S) test applied to the semimajor axis distribution (panel a) gives only a 1.3\% 
probability that the biased model and observed distributions are drawn from the same underlying 
distribution.} 
%The K-S test probabilities for the eccentricity, inclination and absolute magnitude distributions are 
%39\%, 22\% and 48\%, respectively.}
\label{galaxy}
\end{figure}

Our immediate objective in this work is to resolve a longstanding problem with {\it detached} SDOs.
The problem is illustrated in Fig. \ref{galaxy}, where the biased model shows a radial 
profile with far fewer detached SDOs at larger orbital radii than observations.\footnote{We previously found 
-- and confirm it here -- that the population of scattering SDOs do not have the 
same problem. This population is relatively easy to model because its radial profile is practically 
independent of the early orbital evolution of Neptune (see Kaib et al. 2019 for a detailed analysis of 
the inclination distribution of scattering SDOs and Centaurs). Indeed, even a simple model where Neptune's 
orbit is fixed at 30.1 au (i.e., no migration) reproduce the radial profile of the scattering population 
reasonably well.} Something is clearly off here. The 
radial profile problem exists for all migration models that we tested so far. Specifically, we 
tested models with: (1) different timescales of Neptune's migration ($\tau = 5$, 10, 30 and 100 Myr), (2) 
different excitation of Neptune's eccentricity  (e.g., $e_{\rm N}=0$, 0.03 and 0.1), (3) different 
damping timescales of Neptune's eccentricity (e-fold $\tau_e = 5$, 10, 30 and 100 Myr), (4) different 
radial profiles of the original planetesimal disk (Nesvorn\'y et al. 2020), and (5) with and without 
Neptune's jump during the instability. The problem persists independently of 
whether the galactic potential/stars are included in the model (Nesvorn\'y et al. 2017). Moreover, 
while the problem was originally identified when the model was compared with the Outer Solar System 
Origins Survey (OSSOS) detections, it is even more evident in the comparison with the Dark Energy Survey 
(DES) observations of SDOs (Bernardinelli et al. 2022) (Fig. \ref{galaxy}). 

We now believe to have found an interesting solution to this problem. To test it, we performed new 
simulations with a stellar cluster (Adams 2010). The cluster potential and encounters with cluster stars 
were modeled following the methods described in Batygin et al. (2020) (Sect. 2). In all other aspects 
the dynamical model remained the same. As we discuss in Sect. 3, the new model produced the same orbital 
distribution of KBOs for $a<50$ au. For $a>50$ au, however, the model population of detached SDOs is more radially 
extended when the cluster effects are accounted for. Moreover, when we assume that a particularly close 
stellar encounter occurred early on, the model is able to accurately match the radial profile of 
detached SDOs detected by DES (Bernardinelli et al. 2022; Sect. 3). 

The stellar cluster effects were previously invoked to explain extreme KBOs (Kenyon \& Bromley 2004; 
Morbidelli \& Levison 2004; Brasser et al. 2006, 2012; Brasser \& Schwamb 2015), such as (90377) Sedna and 
2012 VP113 ($a>200$ au, $q \simeq 75$--80 au; Brown et al. 2004, Trujillo \& Sheppard 2014). The results 
indicate that the Sun was born in a cluster with $N \sim 10^3$ or more stars, and that the Sun remained 
in the cluster for at least $T \sim 10$ Myr. With only a few extreme KBOs known, however, these results 
are subject to small number statistics. In addition, the extreme KBOs were detected in different 
observational programs that employed different search strategies and had different limiting 
magnitudes. It is therefore not obvious how to model the strong biases involved in their detection.
That is why the radial structure of detached SDOs with $35<q<50$ au, which is well characterized from 
DES observations ($\simeq200$ detected KBOs with $a>50$ au), can represent a useful constraint on 
cluster properties.     

\section{Methods}
  
{\it Migration model.} The numerical integrations consist of tracking the orbits of the 
four giant planets (Jupiter to Neptune) and a large number of planetesimals. Uranus and Neptune are 
initially placed inside of their current orbits and are migrated outward. The {\tt swift\_rmvs4} 
code, part of the {\it Swift} $N$-body integration package (Levison \& Duncan 1994), is used to follow 
all orbits. The code was modified to include artificial forces that mimic the radial migration and 
damping of planetary orbits. The migration histories of planets are informed by our best models of 
planetary migration/instability. Specifically, we adopt the migration model s10/30j from Nesvorn\'y 
et al. (2020) that worked well to satisfy various constraints; see that work for a detailed description 
of the migration parameters (e.g., $\tau=10$ Myr for $t<10$ Myr and instability at $t=10$ Myr). The 
code accounts for the jitter that Neptune's orbit experiences due to close encounters with very 
massive bodies (Nesvorn\'y \& Vokrouhlick\'y 2016). 

{\it Planetesimal Disk.} Each simulations includes one million disk planetesimals distributed from 
4 au to beyond 30 au. Such a high resolution is needed to obtain good statistics for SDOs. We tested 
different initial disk profiles that produced the best fits to the classical Kuiper Belt in 
Nesvorn\'y et al. (2020). For the truncated power-law profile (Gomes et al. 2004), the step in the 
surface density at 30 au is parameterized by the contrast parameter $c \sim 10^3$, which is simply 
the ratio of surface densities on either side of 30 au. The exponential disk profile is parameterized 
by one e-fold $\Delta r \sim 2.5$ au (Nesvorn\'y et al. 2020). The initial eccentricities and inclinations 
of orbits are set according to the Rayleigh distribution. The disk bodies are assumed to be 
massless such that their gravity does not interfere with the migration/damping routines. 

\begin{figure}
\epsscale{0.5}
%\plotone{starenc.eps}
\plotone{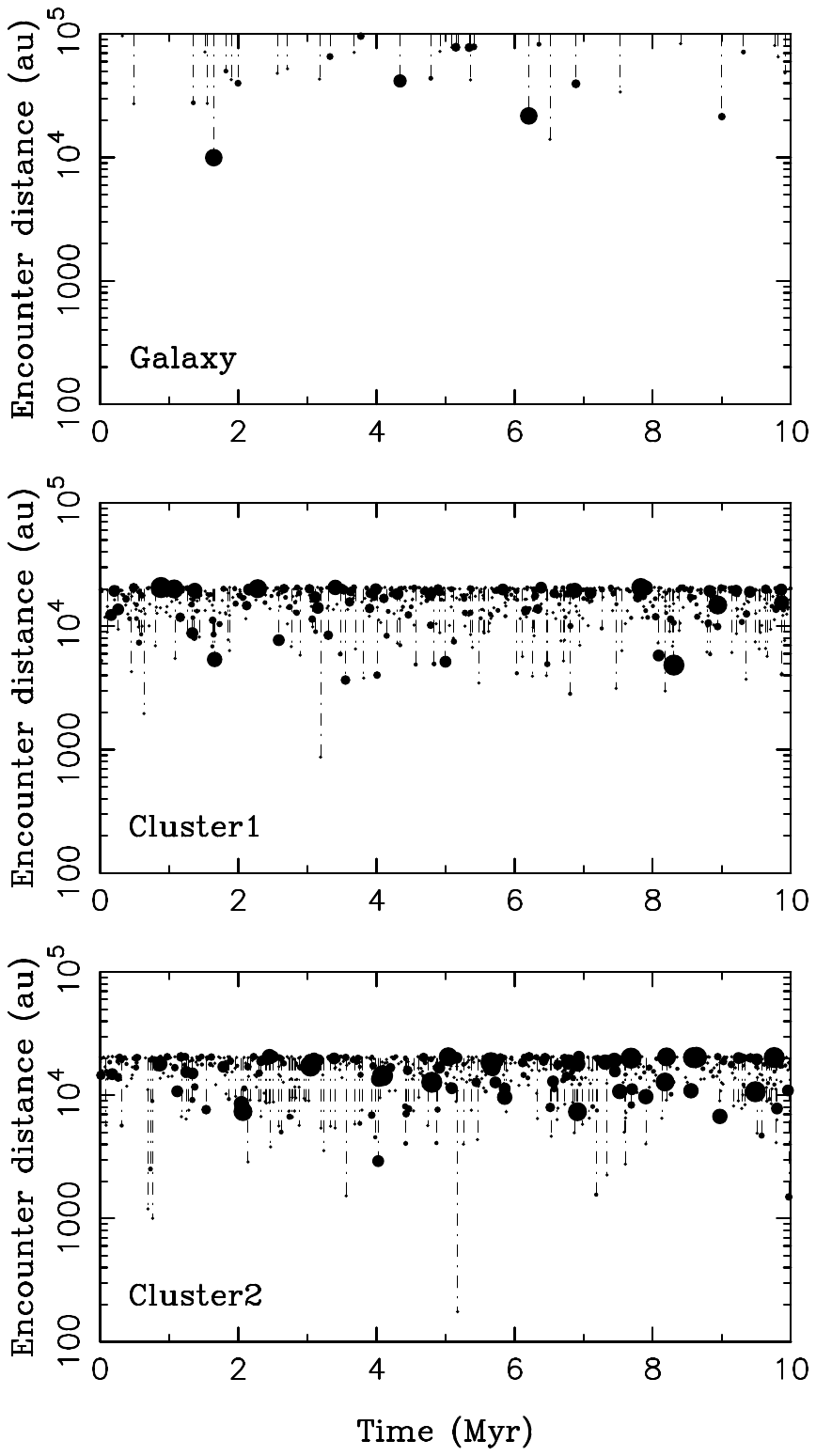}
% gr.starenc.f
\caption{The stellar encounters in the first 10 Myr. From top to bottom, the panels show encounters in
three cases considered in this work: (1) no cluster, encounters of the Sun with stars in the Galaxy
(labeled Galaxy), (2) a cluster with relatively distant stellar encounters (Cluster1), and (3) a 
cluster with relatively close stellar encounters (Cluster2). The size of a symbol correlates with
the stellar mass. The closest encounter for Cluster2 happens at $t\simeq 5.17$ Myr when a 0.17 $M_\odot$ star 
passes at the minimum distance $d \simeq 175$ au. The effects of this encounter on the inclination 
distribution of KBOs and planetary orbits are discussed in Section 4.}
\label{starenc}
\end{figure}

{\it Cluster potential and cluster star encounters.}
The gravitational potential of a cluster (stars and gas) is modeled by the Plummer model (Plummer 1915). 
Adopting the mean stellar mass of $\langle M_* \rangle=0.38$ $M_\odot$ (Kroupa 2001), for the reference 
cluster mass $M = 1200$ $M_\odot$ (roughly comparable to the Orion Nebular Cluster) and the Plummer radius 
$r_{\rm P}=0.35$ pc, the average and central stellar number densities are $\eta  = 100$ pc$^{-3}$ and 
$\eta_{\rm c}=1.7 \times 10^4$ pc$^{-3}$, respectively (Hillenbrand \& Hartmann 1998). We perform two 
simulations for clusters with $N=1000$ stars and the Sun's residence time in the cluster $T=10$ Myr 
(time measured after the gas disk dispersal; $t=0$ in our simulations) that differ in the history of 
stellar encounters with the Sun. [For brevity, we sometimes refer to $T$ as the ``cluster lifetime'', 
but see the discussion below.] In the first case, we opt to model a case 
where all stellar encounters were relatively distant ($r \gtrsim 1000$ au; hereafter the Cluster1 simulation).
In the second case, we model close stellar encounters (Fig. \ref{starenc}; Cluster2). For reference,
we also run an additional case {\it without} the star cluster (Galaxy). The Sun has a orbit 
near the reference radius $\sqrt{2/3}\, r_{\rm P}$. 

Before advancing our discussion, it is imperative to clarify the limitations of the cluster model 
we have adopted. Although the chosen parameters are roughly comparable to the characteristics of 
the Orion Nebula Cluster (ONC), the Plummer model is, by nature, an oversimplification. In a more 
precise rendition of stellar dynamics, the Sun would not maintain a static orbit around the cluster 
core but would instead execute a complex and chaotic trajectory. Furthermore, star clusters themselves 
evolve, with both the gas density and the stellar number density diminishing in time. Detailed modeling 
of these effects would introduce an element of time-dependence into our picture that we currently 
disregard. Nevertheless, we do not expect that these assumptions pose a significant limitation for 
our work, because, as we demonstrate below, our results depend most strongly on the time-integrated 
stellar number density in the Sun's vicinity, not the instantaneous value of $\eta$ itself. 
Consequently, our model can be seen as a means to replace a stochastic integrand with a representative 
average value.

It is also worth highlighting that the quoted cluster lifetime $T$ should not be mistaken for the actual
longevity of the star cluster. It would be more accurate to interpret this period as representing 
the Sun's duration of residence within its birth association (the ONC itself is likely to progress 
into an open cluster over time, possibly bearing resemblance to the loosely-bound Pleiades cluster 
in roughly 100 Myr; Kroupa 2001). Consequently, while our model does not fully 
capture the intricacies of star cluster dynamics, it should provide us with a reliable means to 
investigate the effects of the solar system’s birth environment on the trans-Neptunian region.
Additional constraints on the solar system’s birth environment are discussed in Section 5. 

The effect of stellar encounters is modeled in {\tt swift\_rmvs4} by adding a star at the beginning of its 
encounter with the Sun and removing it after the encounter is over. The stars are released and removed at 
the heliocentric distance of 0.1 pc (20,600 au; increasing this value does not appreciably change the results). 
We use the model of Heisler et al. (1987) to generate stellar encounters but omit white dwarfs to approximate 
the Initial Mass Function (IMF, Kroupa 2001).\footnote{The stars within a birth cluster should be 
close to the IMF, and there should be a larger share of high-mass stars than seen in the galactic field
(Kroupa 2001, Heisler et al. 1987). Given that the results presented here are dominated by the closest 
encounter it seems unlikely that an enhancement in the number density of massive stars would make a 
tangible difference - for close encounters, the results would still be dominated by low-mass stars.}
In contrast to Heisler et al. (1987), we assume a common velocity 
dispersion $\langle v \rangle \sim  1$ km s$^{-1}$ and draw velocities from the Maxwell--Boltzmann distribution 
with a scale parameter $\sqrt{2} \langle v \rangle$ (Binney \& Tremaine 1987). This choice is motivated 
by observational surveys of clusters (Lada \& Lada 2003). 

{\it Galactic potential and stellar encounters.} Effects of the Galaxy become important after the Sun 
leaves the cluster. We assume that the Galaxy is axisymmetric and the Sun follows 
a circular orbit in the Galactic midplane (Sun's migration in the Galaxy is not included; Kaib et al. 2011).
The Galactic tidal acceleration is taken from Levison et al. (2001) (see also Heisler \& Tremaine 1986, Wiegert 
\& Tremaine 1999). The mass density in the solar neighborhood is set to $\rho_0=0.15$ $M_\odot$ pc$^{-3}$.  
The stellar mass and number density of different stellar species are computed from Heisler et al. 
(1987). The stars are released and removed at the heliocentric distance of 1 pc (206,000 au).  For each 
species, the velocity distribution is approximated by the isotropic Maxwell--Boltzmann distribution. 
The dynamical effect of passing molecular clouds is ignored.

{\it Comparison with observations.} 
We compare the model results with DES detections of SDOs. DES covered a contiguous 5000 $\deg^2$ of the 
southern sky between 2013-2019, with the majority of the imaged area being at high ecliptic latitudes. The 
search for outer Solar System objects yielded 812 KBOs with well-characterized discovery biases, 
including over 200 SDOs with $a > 50$ au. The DES observations are more constraining in this work 
than OSSOS given that DES detected more SDOs, as expected from the differences in the geometry 
of both surveys. The DES survey simulator\footnote{Publicly available on \texttt{GitHub} - 
\url{https://github.com/bernardinelli/DESTNOSIM}} (Bernardinelli et al. 2022) enables comparisons between population 
models and the DES data by simulating the discoverability conditions of each member of the test 
population, that is, the model is biased in the same way as the data. These simulations enable the 
application of standard statistical tests (e.g., Kolmogorov–Smirnov) to establish whether a tested 
model can be ruled out from DES observations.

{\it Absolute magnitude distribution.}
After experimenting with different magnitude distributions, we found a setup that works pretty well (Sect. 3). 
The size distribution is assumed to be a broken power law with the knee $D_{\rm knee}=100$ km and the ankle 
$D_{\rm ankle}=300$ km.\footnote{We tested many different possibilities and found that the radial profile of 
detected SDOs is not particularly sensitive to the assumed magnitude distribution (within reasonable limits). 
Specifically, the radial profile of detected SDOs remains practically the same for $D_{\rm knee}=150$ km 
(Lawler et al. 2018).} The size distribution of small bodies with $D<D_{\rm knee}$ is approximated by the 
cumulative power law $N(D) \propto D^{-q_{\rm small}}$ with $q_{\rm small}=2.1$ (Nesvorn\'y 2018). The distant 
KBOs below some minimum size are not detected by DES. For detached SDOs with $60<a<500$ au, we set the 
minimum diameter $D_{\rm min}=70$ km ($D_{\rm min}=40$ km is used for Hot Classicals). The intermediate size 
bodies with $D_{\rm knee}<D<D_{\rm ankle}$ are given $q_{\rm inter}=4.5$ and the large bodies with $D>D_{\rm ankle}$ 
are given $q_{\rm inter}=2.0$. We used the albedo $p_{\rm V}=0.05$ to convert diameters to the absolute (visual) 
magnitudes ($H$). As the DES detections are reported in the red filter, we use the red magnitude $H_{\rm r}=H-0.6$. 
The DES selection function (weakly) depends on the color of each object, so we applied the color 
transformations from Section 2.3 of Bernardinelli et al. (2022) to each object. We also assumed that 
the objects have no variability (i.e. a flat or constant light curve).

\section{Results}

\begin{figure}
\epsscale{0.33}
%\plotone{final1.eps}
\plotone{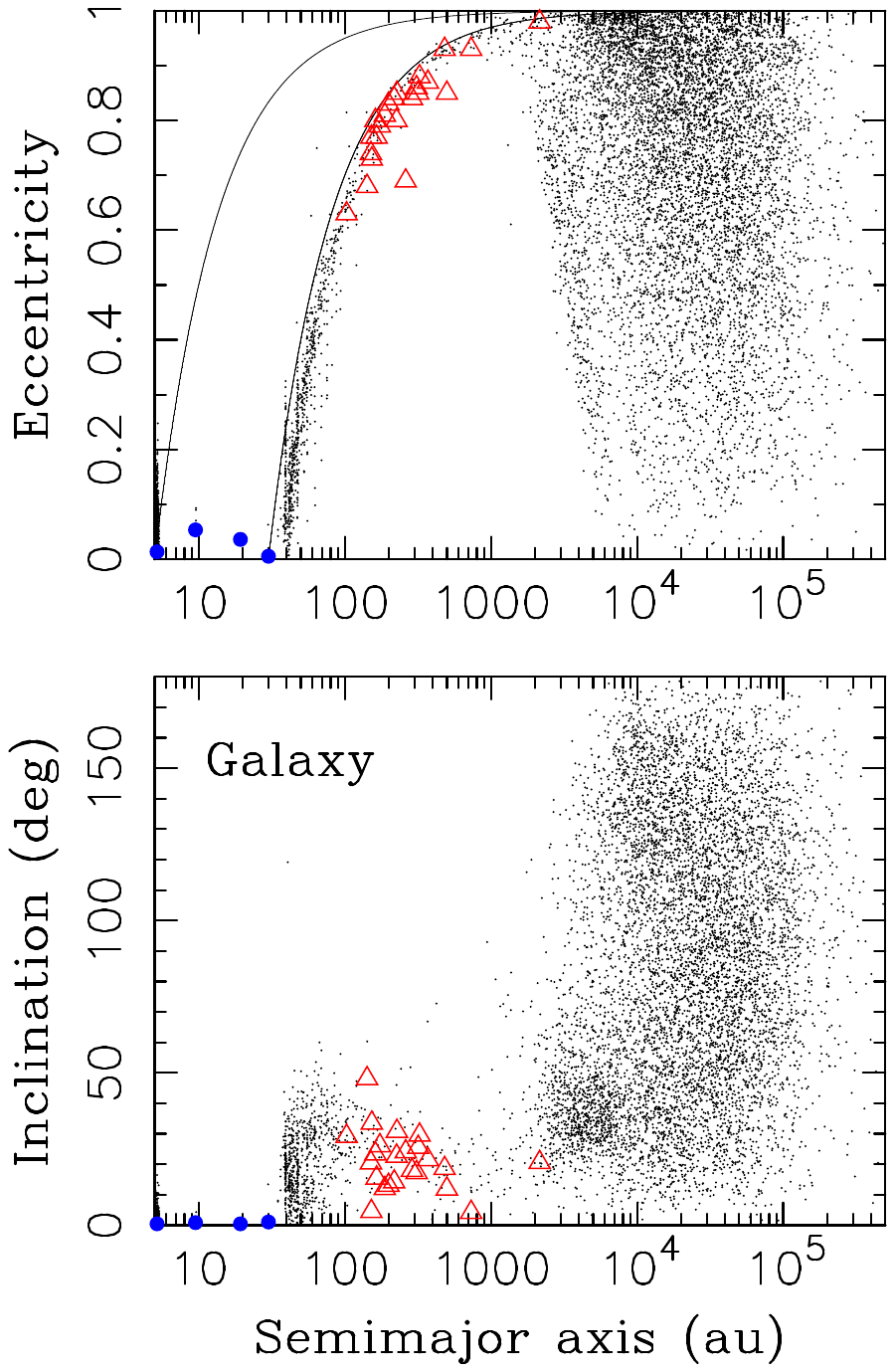}
\epsscale{0.3}
%\plotone{final2.eps}
%\plotone{final3.eps}
\plotone{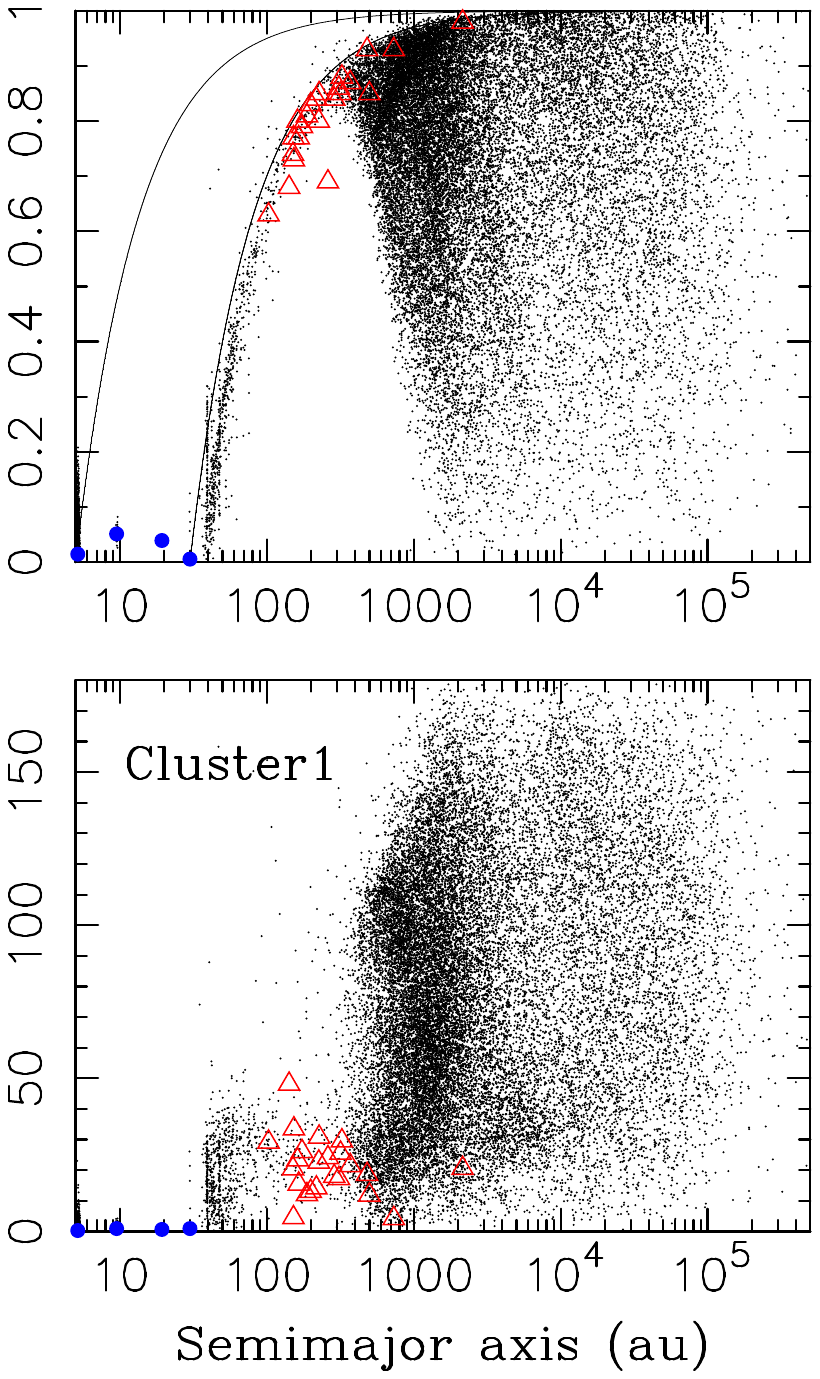}
\plotone{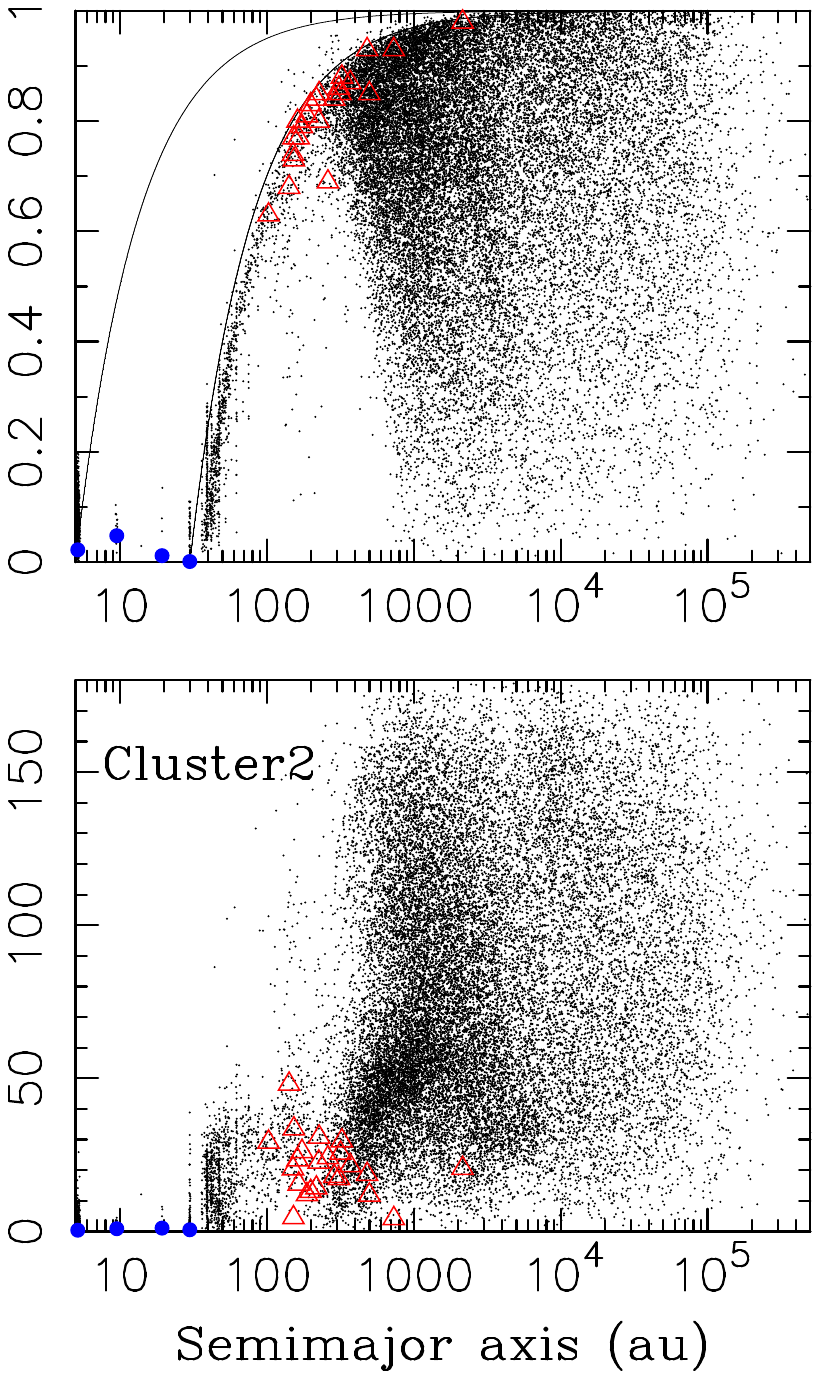}
% gr.galaxy,gr.cluster,gr.cluster2
\caption{The orbital distribution of bodies from three models: (1) no cluster (labeled Galaxy), (2) a cluster with 
relatively distant stellar encounters (Cluster1), and (3) a cluster with relatively close stellar encounters (Cluster2). 
All simulations included the galactic tide and encounters of the Sun with stars in the Galaxy. We sub-sampled the model 
population, shown here at the simulated time $t=4.6$ Gyr (present epoch), by a factor of four to limit saturation. 
The red triangles show orbits of known extreme KBOs.}
\label{final}
\end{figure}

We propose that the problem with the radially extended distribution of detached SDOs (Fig. \ref{galaxy} and 
discussion in Section 1) can be resolved when it is accounted for the effects of close stellar encounters 
during the solar system's cluster stage. To introduce this possibility, we first discuss the results of our 
three simulations -- Galaxy, Cluster1 and Cluster2 -- and point out major differences between them. 
The orbital distributions of bodies obtained in our three models are similar for $a<50$ au and $a>10$,000 au 
but show important differences for $50<a<10$,000 au (Fig. \ref{final}). With the cluster star encounters 
in Cluster1 and Cluster2, bodies with $50<a<10$,000 au can decouple from Neptune and evolve onto orbits with 
lower eccentricities and large inclinations. This creates a spherical cloud of bodies with the overall 
structure similar to the Oort cloud (Oort 1950) but located at smaller orbital radii. We call this the 
Fern\'andez cloud (Fern\'andez 1997). See Fern\'andez (1997), Fern\'andez \& Brunini (2000), Morbidelli
\& Levison 2004, Brasser et al. (2006, 2012) and Kaib \& Quinn (2008) for previous studies.

The boundary between the Oort and Fern\'andez clouds is not well defined. Comparing different panels in Fig. 
\ref{final}, we find that nearly all Oort cloud objects have $a>2$,000 au and the great majority have 
$a>5$,000 au. In our cluster models, the Fern\'andez cloud forms at $a<5$,000 au and extends inwards to 
$a \simeq 200$--300 au. To fix the terminology in this work, the bodies with $250<a<5$,000 are called 
the Fern\'andez cloud objects (FCOs) and the bodies with $a>5$,000 au are Oort cloud objects (OCOs). 
We note that the Oort cloud can divided into two parts: the (active) outer part with $a \gtrsim 15$,000~au 
(Hills 1981, Duncan et al. 1987, Vokrouhlick\'y et al. 2019), which the source of most long-period comets, and 
the (inactive) inner part with $a \lesssim 15$,000 au, where most orbits remain unchanged
(except when very close stellar encounters happen). The outer Oort cloud extends to $a \gtrsim 
10^5$ au.     

Without the stellar cluster, we find that $\simeq3$\% of bodies originally from 4--30 au end up in the 
Oort cloud (Table 1). This fraction does not change when we account for the cluster. This means that 
the stellar cluster environment does not strongly influence the population of OCOs, at least for the 
migration and stellar cluster parameters adopted in this work (Sect.~2, $\tau \sim T$; see Sect.~4 for 
further discussion). With the stellar cluster, roughly 7\% of bodies from 4--30 au end up in the 
Fern\'andez cloud. 

The implantation probabilities given in Table 1 would have to be multiplied by the number of 
planetesimals originally available in each source zone 
to obtain the population estimates for different target zones. For example, for the reference surface 
density profile, $\Sigma \propto 1/r$, there is an equal number of bodies in each semimajor axis interval. 
In this case, the inner SDOs ($50<a<250$ au) and OCOs would predominantly originate from planetesimals 
in the 20--30 au zone, and the Fern\'andez cloud -- for the Cluster1 and Cluster2 models -- would be a 
mix of planetesimals from every zone (implantation probabilities: 6.8--7.2\% from the Jupiter/Saturn zone at 4--10 au,
7.6--8.1\% from the Uranus/Neptune zone at 10--20 au, and 6.3--6.8\% from the outer disk at 20--30 au).

It is expected that the planetesimal populations in the Jupiter/Saturn and Uranus/Neptune 
zones become depleted by the end of the gas disk lifetime ($t=0$ in our simulations).
These planetesimals are scattered inward and outward by the growing giant planets and their orbits 
can be circularized by the gas drag. They can end up the asteroid belt or in the outer disk at $>20$ au 
(Kretke et al. 2012, Raymond \& Izidoro 2018, Vokrouhlick\'y \& Nesvorn\'y 2019). If so, 
the fractions reported for the outer disk -- the last column in Table 1 -- could be the most relevant. 
 
For $\simeq 20$ $M_{\rm Earth}$ of planetesimals between 20 and 30 au (Nesvorn\'y 2018), the Fern\'andez 
and Oort clouds would end up having $\simeq 1.3$ $M_{\rm Earth}$ and $\simeq 1$ $M_{\rm Earth}$, 
respectively (for Cluster2; the estimates for Cluster1 are similar). With $(8 \pm 3)\times10^9$ 
planetesimals in the outer disk with diameters $D>10$ km (Nesvorn\'y et al. 2019), the Fern\'andez and Oort 
clouds would end up having $(5 \pm 2) \times 10^8$ and $(4 \pm 2) \times 10^8$ $D>10$ km bodies 
today.\footnote{If the inner disk below 20 au significantly contributed to the Fern\'andez cloud formation, 
the population and total mass of FCOs could be substantially larger.} The inner SDOs at 50--250 
au should represent a much smaller population with estimated (2.4--$4) \times 10^7$ $D>10$ km 
bodies today. 

\begin{figure}
\epsscale{0.33}
%\plotone{scatter_galaxy.eps}
\plotone{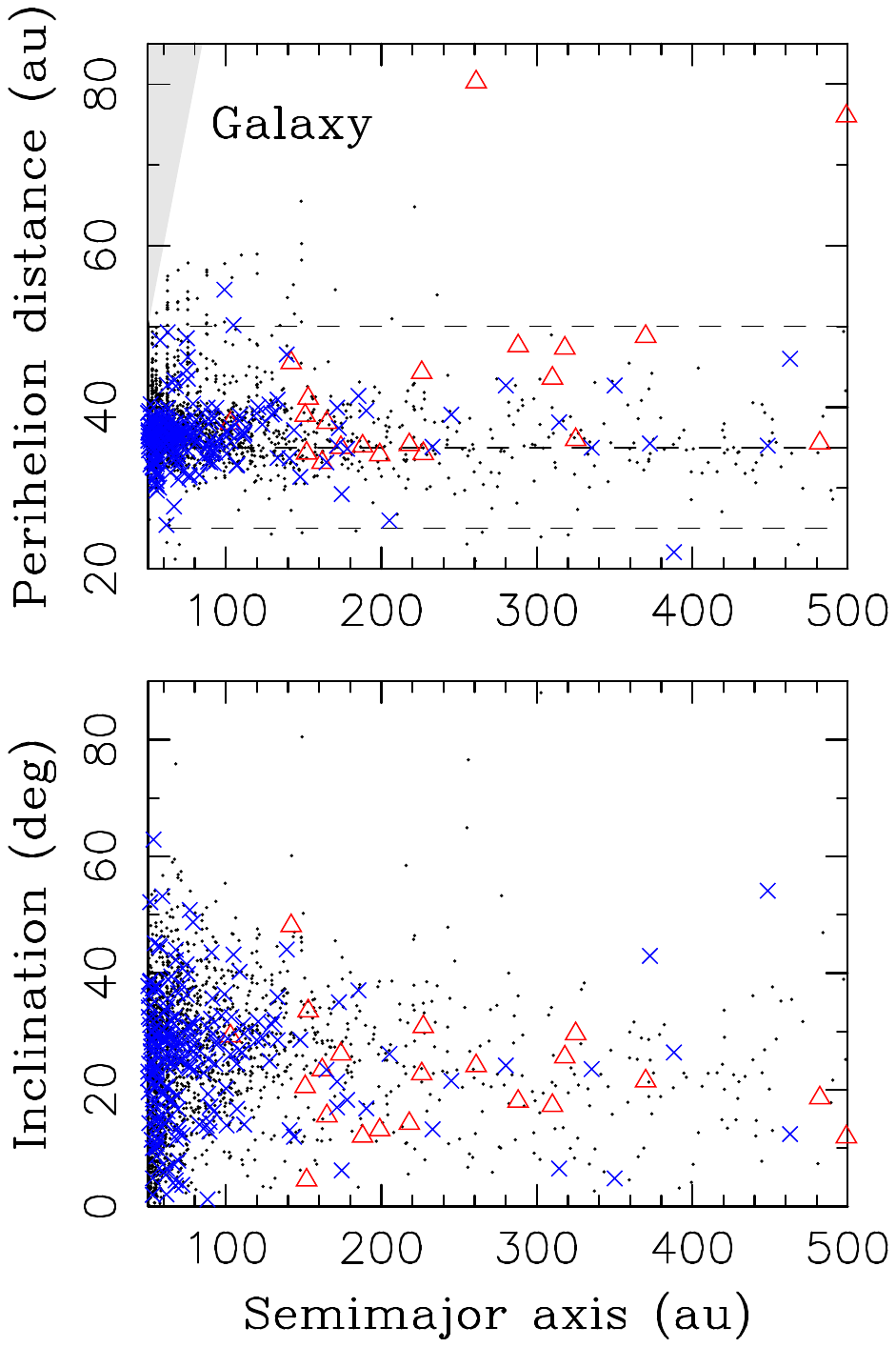}
\epsscale{0.3}
%\plotone{scatter_cluster.eps}
%\plotone{scatter_cluster2.eps}
\plotone{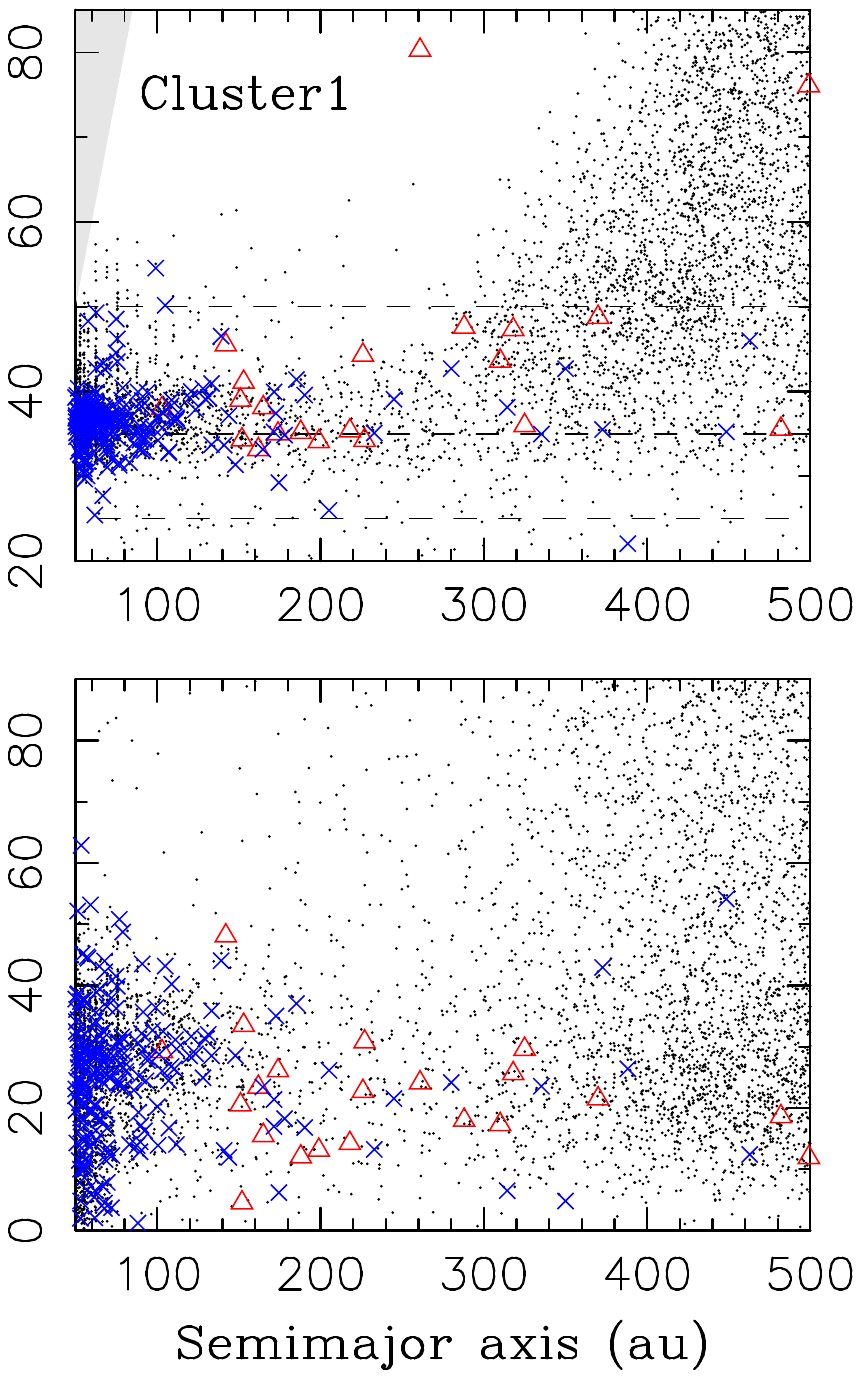}
\plotone{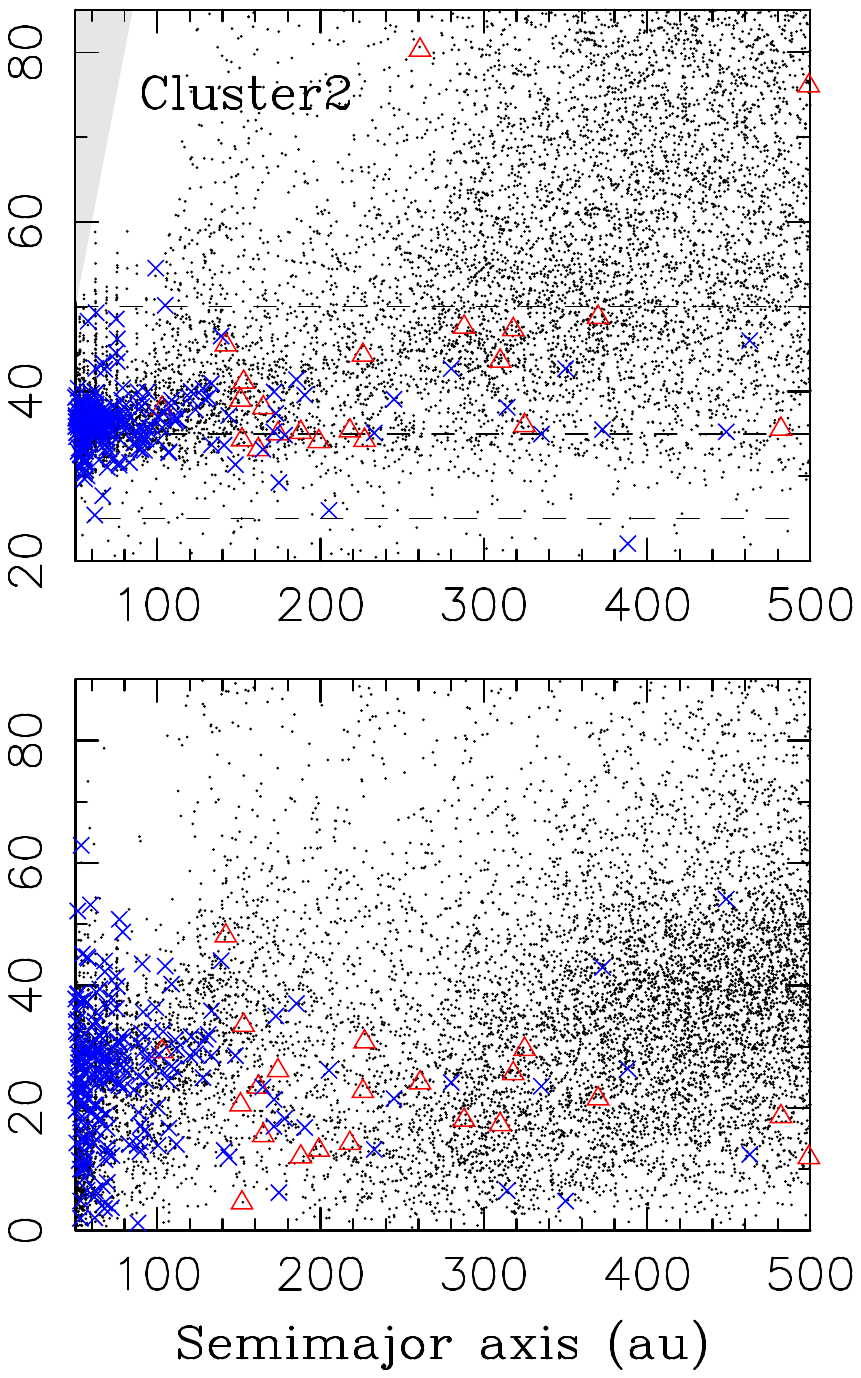}
% gr.scatter_galaxy,gr.scatter_cluster,gr.scatter_cluster2
\caption{The orbital distribution of bodies from three models: (1) no cluster (labeled Galaxy), (2) a cluster with 
relatively distant stellar encounters (Cluster1), and (c) a cluster with relatively close stellar encounters (Cluster2). 
All simulations included the galactic tide and encounters of the Sun with stars in the Galaxy. The model population 
are shown here at the simulated time $t=4.6$ Gyr (present epoch). The red triangles show orbits of known extreme KBOs. 
The blue crosses are DES detections (Bernardinelli et al. 2022).}
\label{scatter}
\end{figure}

\begin{figure}
\epsscale{0.7}
%\plotone{radial1.eps}
%\plotone{radial2.eps}
\plotone{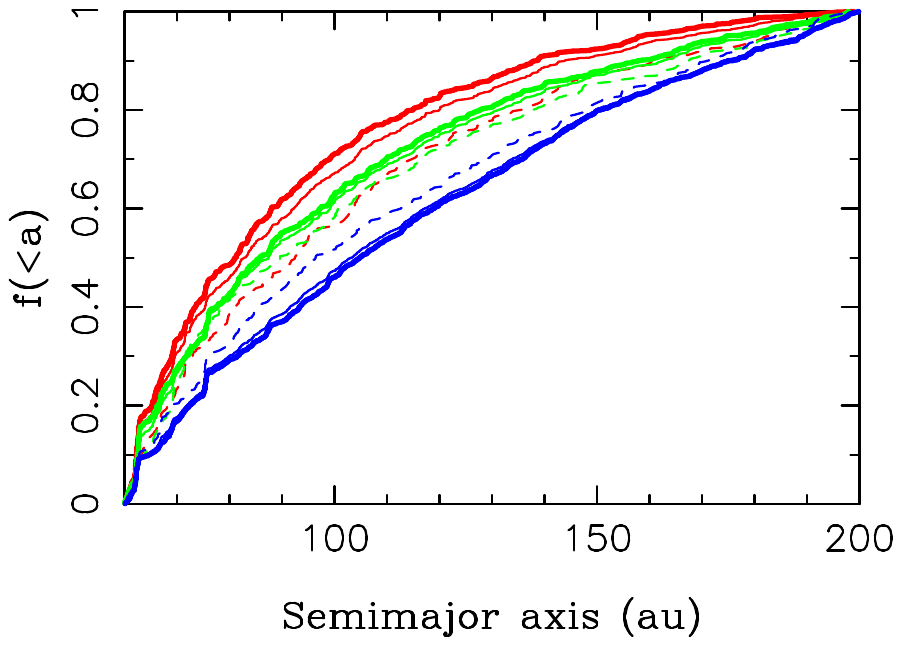}
\caption{The radial profile of bodies in the scattered disk. Different colors show the results from different
simulations: no cluster (red), Cluster1 (green) and Cluster2 (blue). The bold solid lines are the detached populations
with $35<q<50$ au and the thin dashed lines are the scattering disk objects with $25<q<35$ au. The thin solid 
lines are all objects with $25<q<50$ au.}
\label{radial}
\end{figure}

Figure \ref{scatter} shows the orbital distribution for $60<a<500$ au in more detail.  The figure 
highlights the relationship between SDOs and FCOs. In the Galaxy simulation, the detached SDOs with 
$q>35$ au are {\it dropouts} from 
the orbital resonances with Neptune (Kaib \& Sheppard 2016, Lawler et al. 2019). While most 
dropout SDOs have $35<a<50$ au, some with $a \lesssim 150$ au can have higher perihelion 
distances (as high as $q \simeq 60$ au). 
In the Cluster1 and Cluster2 models, FCOs form during close encounters of the cluster stars. Most 
FCOs have orbits with $a>200$ au and $q>50$ au and can be clearly distinguished from the 
dropout SDOs. But there is also a large population of FCOs with $a<200$ au and $35<q<50$ au,
especially in the Cluster2 model, where bodies would be formally classified as detached SDOs.
This shows how the orbital structure of detached SDOs changes when the cluster effects are taken 
into account. Specifically, the radial profile of detached SDOs is more extended in the cluster
simulations than without a cluster (Fig. \ref{radial}). This works in the right direction 
to resolve the problem that motivated this work (Sect. 1).

\begin{figure}
\epsscale{0.8}
%\plotone{cluster.eps}
\plotone{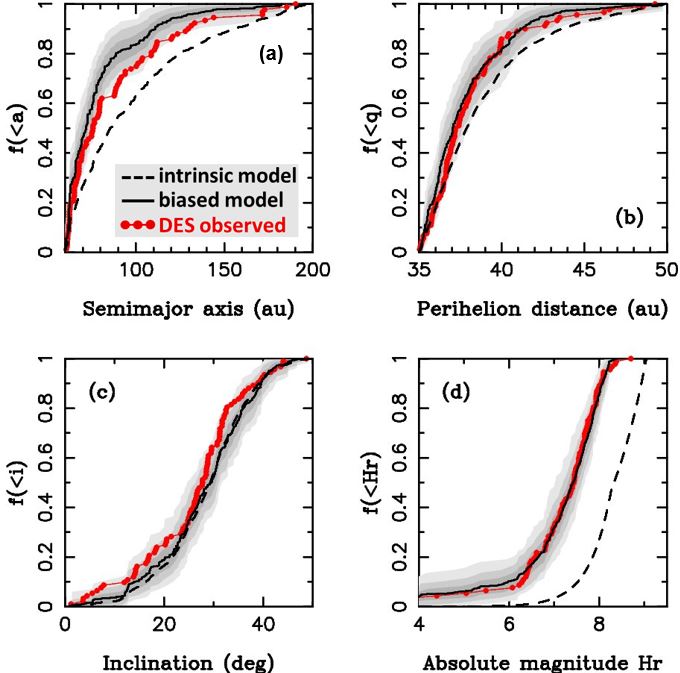}
%scatter8
\caption{A comparison between the biased model (black line) from the Cluster1 simulation and DES observations (red 
lines) of detached SDOs. The intrinsic model distribution is shown as a dashed line. See the caption of Fig. 
\ref{intro} for the meaning of shaded areas. The K-S test applied to the semimajor axis distribution (panel a) gives 
only a 2.9\% probability that the biased model and observed distributions are drawn from the same underlying 
distribution.}
\label{cluster}
\end{figure}

Our Galaxy simulation (i.e., no cluster) was {\it un}successful in reproducing the radial structure of 
detached SDOs observed by DES (Fig. \ref{galaxy}). We applied the Kolmogorov-Smirnov (K-S) test 
to find that the semimajor axis distribution of detached SDOs obtained in this model -- based on 
the comparison of with DES observations -- can be rejected with a 98.7\% probability. Small 
changes of the input size distribution do not significantly influence this result. The comparison 
is done for $60<a<200$ au and $35<q<50$ au because this is where the DES observations are the 
most diagnostic (e.g., only one detached SDO detected by DES with $q>50$ au). The extended 
range $60<a<500$ au is tested below. The radial distribution of detached SDOs obtained in the Cluster1 
model shows a similar problem (Fig. \ref{cluster}). Again, the biased model shows a more compact 
profile than the DES observations. The K-S test applied to the semimajor axis distribution (panel 
a in Fig. \ref{cluster}) suggest that that the model can be rejected with a 97.1\% probability. 

\begin{figure}
\epsscale{0.8}
%\plotone{cluster2.eps}
\plotone{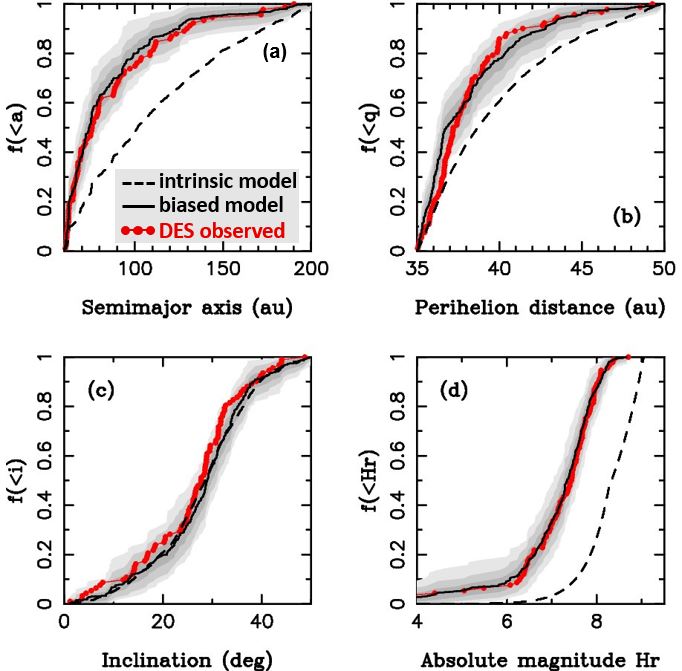}
%scatter6
\caption{A comparison between the biased model (black line) from the Cluster2 simulation and DES observations (red 
lines) of detached SDOs. The intrinsic model distribution is shown as a dashed line. See the caption of Fig. 
\ref{intro} for the meaning of shaded areas. DES detected 92 detached SDOs in the range shown here.}
\label{cluster2}
\end{figure}

\begin{figure}
\epsscale{0.8}
%\plotone{detach500.eps}
\plotone{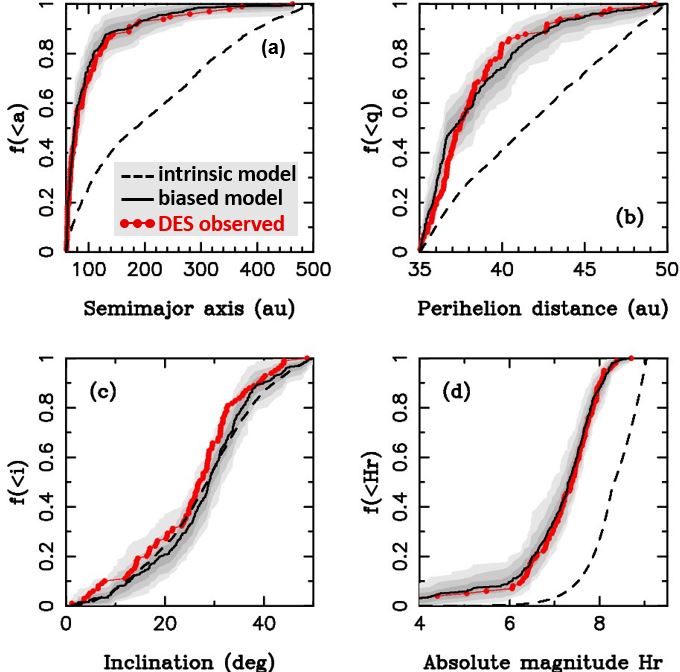}
\caption{The same as Fig. \ref{cluster2} but now for the extended semimajor axis range $60<a<500$ au. See the caption 
of Fig. \ref{intro} for the meaning of lines, symbols and shaded areas. DES detected 99 detached SDOs in the range shown 
here.}
\label{d500}
\end{figure}

\begin{figure}
\epsscale{0.8}
%\plotone{scattering_disk.eps}
\plotone{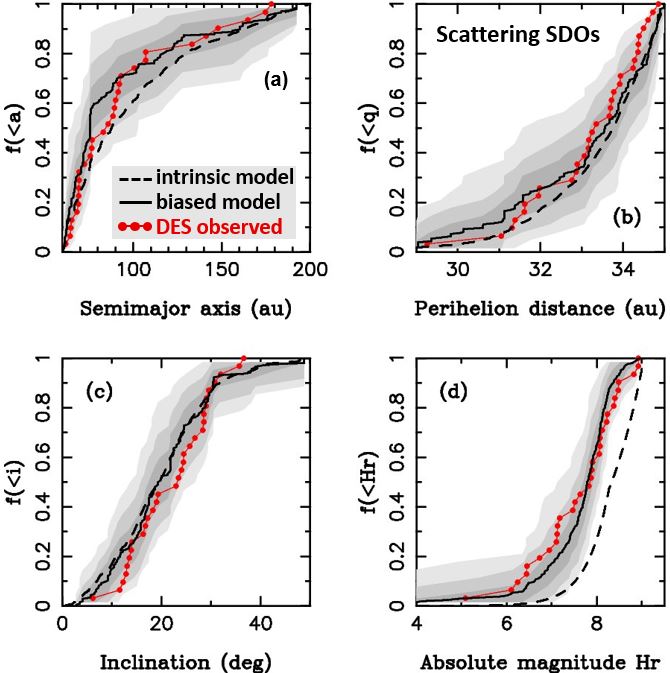}
% gr.scatter2_des.f 
\caption{A comparison between the biased model (black line) from the Cluster2 simulation and DES observations (red 
lines) of scattering SDOs. The intrinsic model distribution is shown as a dashed line. See the caption of Fig. 
\ref{intro} for the meaning of shaded areas. The statistics for the scattering SDOs shown here is as not as good 
as the one shown for the detached SDOs in Fig. \ref{cluster2}, because DES detected only 31 scattering SDOs in the 
range shown here.}
%K-S probabilities: 40\%, 46\%, 26\%, 27\%. }
\label{sdisk}
\end{figure}

Finally, in the Cluster2 model with a very close stellar encounter (0.17 $M_\odot$ star at distance 
$d \simeq 175$ au), the match to DES observations is good (Fig. \ref{cluster2}).\footnote{An important
difference between our two cluster simulations is that Cluster2 shows orbits similar to those of  
(90377) Sedna ($a=506$ AU, $q=76.2$a au). These orbits do not exist in Cluster1. This, in itself, 
can be taken as an argument to favor Cluster2 over Cluster1.} The population of detached 
SDOs is radially extended as it should be, the perihelion and inclination distributions look great 
as well. The K-S test applied to the semimajor axis distribution gives a 84\% probability that
the two distributions are statistically the same. The agreement in the extended semimajor axis 
range is equally good (Fig. \ref{d500}; 82\% K-S test probability). For completeness, 
we also show a comparison between the Cluster2 model and DES observations for the {\it scattering} disk 
(Fig. \ref {sdisk}). As we explained in Sect. 1, the scattering disk population is particularly easy 
to model because its current orbital structure is practically independent of the dynamical evolution 
of the early Solar System.  

For Cluster2, our results suggest that there should be a transition in the orbital structure 
of trans-Neptunian objects (TNOs) near 100 au. For $a \lesssim 100$ au, most TNOs should be dropouts 
from resonances with migrating Neptune (Kaib \& Sheppard 2016, Nesvorn\'y et al. 2016). The detached 
SDOs with $a \lesssim 100$ au should thus concentrate near orbital resonances (e.g., 3:1, 
4:1, 5:1; Lawler et al. 2019). The resonant dropouts have moderate perihelion distances and 
moderate inclinations ($q \lesssim 60$ au, $i \lesssim 50^\circ$). The orbital structure should change for
$a > 100$ au, where most TNOs should have decoupled from Neptune during the cluster stage  
(see Cluster2 in Fig. \ref{scatter}). These FCOs can have large perihelion distances and large 
inclinations, and their number should increase with the semimajor axis (the FCO population
at 250--5000 au can represent several Earth masses). We do not find any diagnostic correlations
between different orbital parameters. There is a slight enhancement of FCO inclinations near 
40$^\circ$, which is probably related to the geometry of the closest encounter in our Cluster2 
simulation. In Cluster1, there is a slight enhancement for $i \simeq 30^\circ$.

\section{Constraints from Cold Classicals}

Above we showed that a close stellar encounter, e.g., 0.17 $M_\odot$ star at distance $d \simeq~175$ au,
could have affected the radial profile of detached SDOs. Here we ask whether such a close encounter would
be compatible with the inclination distribution of Cold Classicals (CCs). CCs are population of KBOs with 
orbits $42 < a < 47$ au, $q > 36$ au and $i < 5^\circ$ (Gladman et al. 2008). They are thought to 
have formed in situ and remained largely undisturbed during Neptune's migration (Batygin et al. 2011,
Dawson \& Murray-Clay 2012).

Batygin et al. (2020) highlighted the importance of the inclination distribution of CCs as an important 
constraint on stellar encounters.\footnote{The inclination distribution of CCs represents a stronger 
constraint on stellar encounters than the stability of planetary orbits (e.g., Adams \& Laughlin 2001).
We verified that the change in planetary orbits from stellar encounters in the Cluster1 and Cluster2
simulations was negligible.} 
The inclination distribution of CCs is well described by a Rayleigh distribution with a scale parameter 
$\sigma_i=1.7$ deg (mean inclination $\langle i \rangle = \sqrt{\pi/2}\, \sigma_i = 2.1$ deg). Batygin et al. 
(2020) showed that the low orbital inclinations of CCs can be used to set an upper limit on $\eta T$, because very 
close stellar encounters could happen for large values of $\eta T$, and these encounters would excessively 
excite inclinations.   

The fraction of cluster realizations that are incompatible with the the inclination distribution 
of CCs increases with $\eta T$. For example, clusters with $\eta T > 3 \times 10^4$ Myr pc$^{-3}$
have a $\gtrsim 50$\% probability to be incompatible (Sect. 5.2 in Batygin et al. 2020).
For $\eta T \sim 10^4$ Myr pc$^{-3}$, which is the reference cluster tested here, the 
probability of being incompatible is only $\sim 20$\%. We therefore see that the cluster 
parameters adopted in this work are consistent with low orbital inclinations of CCs.

\begin{figure}
\epsscale{0.8}
%\plotone{ccs_labeled.JPG}
\plotone{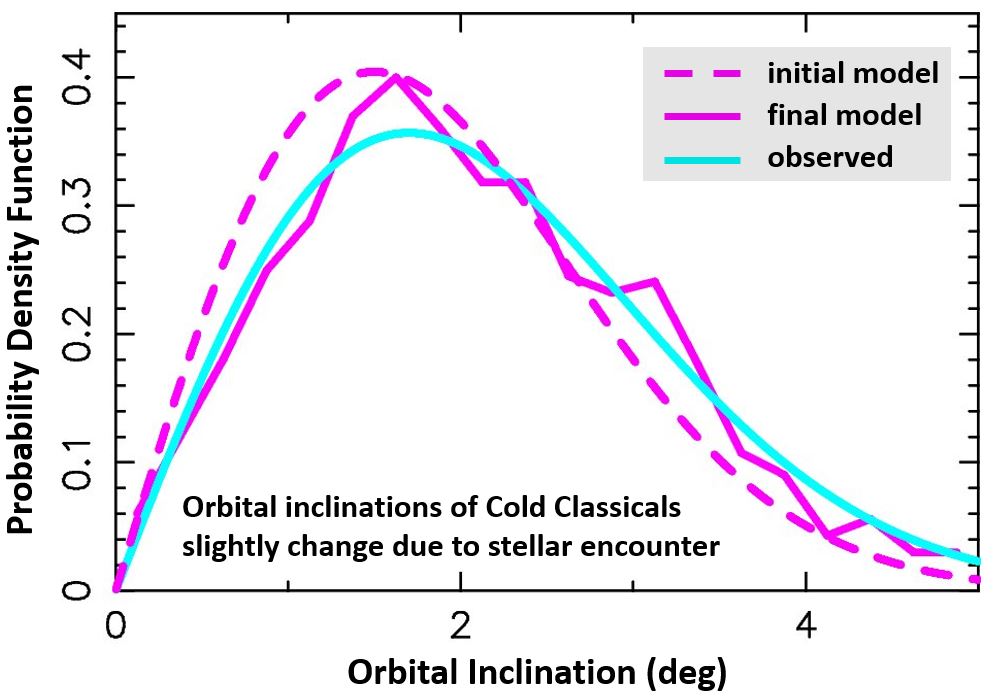}
% gr.inclination.f from encounter_real4 
\caption{The observed inclination distribution of CCs (blue line) is compared to model results 
(purple lines). We distributed 1,000 test CCs on initial orbits with $42<a<47$ au and $q>36$ au. The 
initial inclinations followed a Rayleigh distribution with $\sigma_i=1.5$ deg (dashed line). The orbits 
were integrated for 10 Myr in the Cluster2 model, including the close stellar encounter at $t\simeq 
5.17$ Myr (Fig. \ref{starenc}). The inclination distribution did not change much over the course of 
the integration, and the final distribution of model CCs (solid purple line) is a good match to 
observations.}
\label{ccs}
\end{figure}

To verify this, we performed additional Cluster1 and Cluster2 simulations where test 
bodies were distributed on the initial orbits with $42 < a < 47$ au and $q > 36$ au 
(Figure \ref{ccs}). We tested different initial inclination distributions. The inclination 
excitation of CCs in Cluster2 simulations was found to be $\delta i \sim 1$ deg (there is some 
variability with the geometry of the closest stellar encounter); we did not find much excitation 
for Cluster1. The stellar encounters cannot be the cause of CC inclinations, however, 
because they produce inclination distributions that are much more sharply peaked than the observed 
distribution (Batygin et al. 2020). It thus seems possible that some other process, such 
as dynamical self-stirring of CCs (Batygin et al. 2020), should be responsible for the inclination 
distribution of CCs. 

\section{Discussion}
The results described here suggest that a particularly close stellar encounter happened early 
in the Solar System history. We are not able to characterize the properties of this encounter 
in detail due to the small number of simulations performed in this 
work\footnote{The simulations are computationally expensive: one full simulation for 4.6 Gyr 
requires $\simeq500$ hours on 2000 Ivy Bridge cores of NASA's Pleiades Supercomputer.}. The 
case that works is an encounter of a $\simeq 0.17$ $M_\odot$ star at the distance $d \simeq 175$ au 
from the Sun. The probability of such an encounter is negligible if it is not accounted for 
the Sun's birth environment in a stellar cluster. To have a reasonable probability, the stellar 
cluster must have been sufficiently dense and/or long lived. 

Here we obtained the stellar encounters by modeling a stellar cluster with the average stellar number 
density $\eta  = 100$ pc$^{-3}$ and lifetime $T=10$ Myr. We tested two different cases, one with 
``typical'' stellar encounters for such a cluster (Cluster1) and one with a particularly close encounter
(Cluster2). We now ask how likely it is to have such a close encounter in the tested cluster. To 
estimate the probability, we randomly generate stellar encounters and establish whether at least one 
encounter with $d < 200$ au happens in 10 Myr. This gives the probability of 25\%, and is consistent
with a simple estimate of the rate of stellar encounters based on the $n \sigma v$ argument. 
%Also for second-to-last paragraph: I take \eta = 1.7 e+4 (1+2/3)^{-5/2} = 4740/pc^3 and <v> = \sqrt{2} 
%1 km/s. I get eta * pi * (200AU)^{2} <v> (10 Myr) = 0.2 so I agree with 25\%.

The number of close encounters scales with the product of the stellar number density and residence 
time of the Sun in the cluster, $\chi = \eta T$, with $\chi = 10^4$ Myr pc$^{-3}$ for our nominal cluster. 
Thus, for example, to have a 50\% or larger probability of an encounter with $d < 200$ au, we infer a cluster 
with $\chi > 2.5 \times 10^4$ Myr pc$^{-3}$. For $T=10$ Myr, the cluster would either need to have 
$\gtrsim 2$,500 stars or be more compact ($r_{\rm P} \lesssim 0.14$ pc). Longer Sun's residence times 
in the cluster work as well. For reference, Kobayashi \& Ida (2001) estimated the distance of the 
closest encounter in a stellar cluster as
\begin{equation}
d \sim 200\, {\rm au} \left(N \over 2 \times 10^3\right) \left( 2\,{\rm pc} \over R \right)\ ,  
\end{equation}
where $R$ is the cluster radius. 

Batygin et al. (2020) suggested $\chi \lesssim 3 \times 10^4$ Myr pc$^{-3}$ based on constraints from 
the inclination distribution of cold KBOs (see Sect. 4 above). Together, these results would
indicate $\chi \simeq 1$--$3 \times 10^4$ Myr pc$^{-3}$, which would be a remarkably tight limit 
on the Sun's birth environment. At the same time, the fact that the encounter envisioned here
is near the acceptable limit allowed from the inclination distribution of CC is intriguing, and 
warrants further investigation into the existence of alternative models that may relax this constraint. 

Arakawa \& Kokubo (2023) considered constraints on Sun's cluster properties from direct injection 
of $^{26}$Al-rich materials from a nearby core-collapse supernova. Their results depend on the 
(unknown) duration of star formation, $t_{\rm SF}$. For example, for $t_{\rm SF} \simeq 10$ Myr,
the cluster should have had $N \sim 2000$ stars for at least one core-collapse supernova to 
happen with a 50\% probability. This constraint is thus broadly consistent with the ones 
discussed above.

We assumed that $\tau \sim T$ throughout this work, i.e., that the Neptune's migration timescale is 
similar to the Sun's residence time in the stellar cluster. With this setup, planetesimals are 
scattered outward by the planets when the stellar cluster is still around and this leads to 
the massive Fern\'andez cloud formation ($\gtrsim 1.3$ $M_{\rm Earth}$). At the same time, as the Sun leaves 
the cluster at $t=10$ Myr in our simulations, there is plenty of time after the cluster stage to 
form a sufficiently massive Oort cloud ($\sim 1$ $M_{\rm Earth}$), as needed to explain observations 
of the Oort cloud comets (Vokrouhlick\'y et al. 2019). But what if $\tau \gg T$ or $\tau \ll T$? 

In the first case, for $\tau \gg T$, planetesimals from the Jupiter/Saturn region would still be 
scattered early, within the the cluster lifetime, but those from the outer disk would be delayed
(as Neptune takes time to reach 30 au). This would presumably reduce the mass of the Fern\'andez cloud
and make it more difficult to reproduce the radial profile of detached SDOs. This argument 
suggests that the models with very slow migration of Neptune (instability at $t \gg 10$ Myr) could
be in some tension with DES observations of detached SDOs. In the second case, with $\tau \ll T$,
given that Neptune's migration is thought to have been slow ($\tau \gtrsim 10$ Myr; Nesvorn\'y 
2015a), we would have $T \gg 10$ Myr. Here, planetesimals from the whole source region (4--30 au) 
would be scattered outward during the cluster stage. This would presumably reduce the implantation 
probability of planetesimals into the Oort cloud, relative to the case with $\tau \sim T$, and
could be in conflict with the number and properties of the Oort cloud comets (Vokrouhlick\'y 
et al. 2019). A detailed investigation of these issues is left for future work.   

We have not investigated the possibility that the radial structure of detached SDOs was affected by planet 
9 (Trujillo \& Sheppard 2014, Batygin \& Brown 2016, Kaib et al. 2019) 
or a rogue planet (Gladman \& Chan 2006). Whether or not planet 9 exists will probably be established in 
the near future (Schwamb et al. 2023). If it exists, we will know its mass and orbit,
and this should make it possible to show -- via additional modeling --  whether it could have affected 
the radial distribution of detached SDOs with $a<200$ au. After being scattered by Neptune, the rogue 
planet of Gladman \& Chan (2006) could have had a complex orbital history. A statistically large ensemble 
of simulations will presumably be needed to establish the range of possibilities in this model.   
 
\section{Conclusions}

We pointed out a longstanding problem with the radial distribution of detached SDOs and showed that this 
problem can be resolved if the Sun had a particularly close encounter with a cluster star (e.g., M dwarf 
with the mass $\simeq 0.2$ $M_\odot$ and minimum distance $\lesssim 200$ au). With such a close encounter,
a large population of Fern\'andez cloud objects forms, and this population extends to $a<200$ au and $q<50$ au,
where it affects the radial structure of detached SDOs detected by the Dark Energy Survey (Bernardinelli 
et al. 2022). We performed three new simulations to document this effect in detail. The orbital distributions 
of detached SDOs obtained in different models were biased with the DES simulator to allow for a one-to-one 
comparison with the observations. We also applied the same method to a dozen of our previous models, which 
varied in their assumptions on the properties of Neptune's migration but did not include the effects of the 
stellar cluster, to demonstrate that many possibilities related to the effect of Neptune's migration can 
be ruled out. The investigation of the effects of planet 9 or a rogue planet in the radial structure of 
the scattered disk is left for future work. 

\acknowledgements
The work of DN was supported by the NASA Emerging Worlds program. PHB acknowledges support from the 
DIRAC Institute in the Department of Astronomy at the University of Washington. The DIRAC Institute is 
supported through generous gifts from the Charles and Lisa Simonyi Fund for Arts and Sciences, and the 
Washington Research Foundation. The work of DV was supported by the Czech Science Foundation 
(grant number 21--11058S). We thank Matt Clement and an anonymous reviewer for helpful comments
on the submitted manuscript.

\begin{table}
\centering
{
\begin{tabular}{lrrrrr}
\hline \hline   
 & source & whole range  & J/S zone  & U/N zone  & outer disk    \\
 &        &     4--30 au & 4--10 au         & 10--20 au       & 20--30 au     \\
target    &     & \% & \% & \% & \%  \\   
\hline
& & \multicolumn{4}{c}{\it Galaxy} \\
scattered disk  & 50--250 au          & 0.17   & 0.02   & 0.11    & 0.31  \\ 
Fern\'andez cloud    & 250--5000 au       & 0.39   & 0.06   & 0.32    & 0.66  \\
inner Oort      & 5000--15,000 au     & 0.93   & 0.30   & 0.81    & 1.4   \\ 
outer Oort      & 15,000--200,000 au  & 2.0    & 1.0    & 1.8     & 2.7   \\
\hline
& & \multicolumn{4}{c}{\it Cluster1} \\
scattered disk  &50--250 au         & 0.19   & 0.03   & 0.14    & 0.33  \\ 
Fern\'andez cloud     &250--5000 au       & 7.4    & 7.2    & 8.1     & 6.8   \\
inner Oort      &5000--15,000 au     & 1.3    & 0.47   & 1.2     & 1.9   \\ 
outer Oort      &15,000--200,000 au  & 1.6    & 0.39   & 1.4     & 2.5   \\ 
\hline
& & \multicolumn{4}{c}{\it Cluster2} \\
scattered disk &50--250 au         & 0.34   & 0.11   & 0.32    & 0.5   \\ 
Fern\'andez cloud    &250--5000 au       & 6.9    & 6.8    & 7.6     & 6.3   \\
inner Oort     &5000--15,000 au     & 1.5    & 0.65   & 1.4     & 2.1   \\ 
outer Oort     &15,000--200,000 au  & 1.7    & 0.45   & 1.5     & 2.6   \\ 
\hline \hline
\end{tabular}
}
\caption{The implantation probability for different source and target regions.
This is the probability that a body starting in the source region at $t=0$ 
(the gas disk dispersal) ends up in the target region at $t=4.6$ Gyr (present epoch).}
\end{table}

\clearpage

\clearpage

\clearpage

\clearpage

\clearpage

\clearpage

\clearpage

\clearpage

\clearpage

\clearpage

\clearpage

\end{document}